\documentclass[aps,prb,twocolumn,superscriptaddress]{revtex4}
\usepackage{amsmath,amsfonts,amssymb}
\usepackage{graphicx,float,calc}
\usepackage{color,bm}
\usepackage{ulem}
\usepackage{braket}
\usepackage{hyperref}
\begin{document}

\title{Skin superfluid, topological Mott insulators, and asymmetric dynamics in an interacting non-Hermitian Aubry-Andr\'{e}-Harper model}

\author{Dan-Wei Zhang}
\email{danweizhang@m.scnu.edu.cn}
\affiliation{Guangdong Provincial Key Laboratory of Quantum Engineering and Quantum Materials, GPETR Center for Quantum Precision Measurement and SPTE, South China Normal University, Guangzhou 510006, China}
\affiliation{Frontier Research Institute for Physics,
South China Normal University, Guangzhou 510006, China}

\author{Yu-Lian Chen}
\affiliation{Guangdong Provincial Key Laboratory of Quantum Engineering and Quantum Materials, GPETR Center for Quantum Precision Measurement and SPTE, South China Normal University, Guangzhou 510006, China}

\author{Guo-Qing Zhang}\email{zhangptnoone@m.scnu.edu.cn}
\affiliation{Guangdong Provincial Key Laboratory of Quantum Engineering and Quantum Materials, GPETR Center for Quantum Precision Measurement and SPTE, South China Normal University, Guangzhou 510006, China}
\affiliation{Frontier Research Institute for Physics,
South China Normal University, Guangzhou 510006, China}

\author{Li-Jun Lang}
\affiliation{Guangdong Provincial Key Laboratory of Quantum Engineering and Quantum Materials, GPETR Center for Quantum Precision Measurement and SPTE, South China Normal University, Guangzhou 510006, China}

\author{Zhi Li}
\affiliation{Guangdong Provincial Key Laboratory of Quantum Engineering and Quantum Materials, GPETR Center for Quantum Precision Measurement and SPTE, South China Normal University, Guangzhou 510006, China}

\author{Shi-Liang Zhu}\email{slzhu@nju.edu.cn}
\affiliation{National Laboratory of Solid State Microstructures and School of Physics, Nanjing University, Nanjing 210093, China}
\affiliation{Guangdong Provincial Key Laboratory of Quantum Engineering and Quantum Materials, GPETR Center for Quantum Precision Measurement and SPTE, South China Normal University, Guangzhou 510006, China}
\affiliation{Frontier Research Institute for Physics,
South China Normal University, Guangzhou 510006, China}

\begin{abstract}

Non-Hermitian quantum many-body systems are a fascinating subject to be explored. Using the generalized density matrix renormalisation group method and complementary exact diagonalization, we elucidate the many-body ground states and dynamics of a 1D interacting non-Hermitian Aubry-Andr\'{e}-Harper model for bosons. We find %well-defined and
stable ground states in the superfluid and Mott insulating regimes under wide range of conditions in this model. We reveal a skin superfluid state induced by the non-Hermiticity from the nonreciprocal hopping. We investigate the topology of the Mott insulating phase and find its independence of the non-Hermiticity. The topological Mott insulators in this non-Hermitian system are characterized by four equal Chern numbers and a quantized shift of biorthogonal many-body polarizations. Furthermore, we show generic asymmetric expansion and correlation dynamics in the system. %Our developed numerical methods can be used to explore non-Hermitian many-body physics in both equilibrium and nonequilibrium cases.

\end{abstract}

\date{\today}

\maketitle

\section{Introduction}
Non-Hermitian systems have intriguing physics and applications beyond Hermitian systems \cite{Bender1998,LFeng2017,El-Ganainy2018,Miri2019}. Recently, the topological phases in noninteracting non-Hermitian systems have been widely studied \cite{Rudner2009,Zeuner2015,Esaki2011,YHu2011,BZhu2014,Malzard2015,Mejia-Cortes2015,Leykam2017,YXu2017b,XZhan2017,Lee2016,SYao2018,SYao2018b,FSong2019,Kunst2018,YXiong2018,LJin2019,Borgnia2019,ZGong2018,HShen2018,Takata2018,YChen2018,LJLang2018,YWang2019,Harari2018,Bandres2018,HZhou2018,TSDeng2019,Ezawa2019,XWLuo2019b,Kawabata2019,Ghatak2019,Kawabata2018,JQCai2018,TLiu2019,Lee2019,Longhi2019,HJiang2019,QBZeng2019,JHou2019,DWZhang2020,XWLuo2019,Alvarez2018}.
In addition, non-Hermitian many-body physics is expected to be a fascinating but much less explored area \cite{Kim2001,Carmele2015,Levi2016,Tripathi2016,Ashida2017,Nakagawa2018,Lourenco2018,Hamazaki2018,Yamamoto2019,Okuma2019,WXi2019,Lee2019a,Yoshida2019,Luitz2019,SMu2019,CXGuo2019,Matsumoto2019}.
Notably, the interplay between non-Hermiticity and interactions can bring exotic quantum many-body effects, such as non-Hermitian extensions of Kondo effect \cite{Nakagawa2018,Lourenco2018}, many-body localization \cite{Hamazaki2018}, and fermionic superfluidity \cite{Yamamoto2019,Okuma2019}. Topological states in one-dimensional (1D) interacting non-Hermitian Su-Schrieffer-Heeger model \cite{WPSu1979} and fractional quantum Hall system with non-Hermitian interactions are revealed \cite{WXi2019,Lee2019a,Yoshida2019}. However, most of these work focus on the static properties in the mean-field regime or small systems with few particles \cite{Kim2001,Carmele2015,Levi2016,Tripathi2016,Ashida2017,Nakagawa2018,Lourenco2018,Hamazaki2018,Yamamoto2019,Okuma2019,Yoshida2019,Luitz2019,WXi2019,Lee2019a,SMu2019,CXGuo2019,Matsumoto2019}. Interacting non-Hermitian systems of large sizes and their dynamics remain largely unexplored, which is partially due to the lack of efficient numerical tools for non-Hermitian quantum many-body systems. For instance, the density matrix renormalisation group (DMRG) \cite{White1992,schollwock2011density} is one of the most powerful numerical methods for 1D strongly correlated Hermitian systems, but the convergence of the calculation is not guaranteed for a non-Hermitian Hamiltonian.

In this paper, based on our generalized DMRG method and complementary exact diagonalization (ED), we elucidate the many-body ground states and quantum dynamics of a 1D interacting non-Hermitian Aubry-Andr\'{e}-Harper (AAH) model \cite{Harper1955,Aubry1980} for bosons. This model has not been studied in previous work \cite{Kim2001,Carmele2015,Levi2016,Tripathi2016,Ashida2017,Nakagawa2018,Lourenco2018,Hamazaki2018,Yamamoto2019,Okuma2019,Yoshida2019,Luitz2019,WXi2019,Lee2019a,SMu2019,CXGuo2019,Matsumoto2019},
probably because the usually used DMRG fails to study this model in the non-Hermitian case. We here first improve the DMRG approach and then investigate the many-body non-Hermitian AAH model and our main results for this model are as follows: (i) We uncover well-defined and stable ground states in the superfluid and Mott insulating phases in this non-Hermitian system under both open boundary conditions (OBCs) and periodic boundary conditions (PBCs). (ii) We reveal a skin superfluid state under OBCs induced by non-Hermiticity from the nonreciprocal hopping. (iii) We investigate the topological properties of the Mott insulating phase and find that the topological Mott insulators (TMIs) \cite{Raghu2008,SLZhu2013,XDeng2014,Kuno2017,Grusdt2013,YLChen2020} independence of the non-Hermiticity. The TMIs in our non-Hermitian system are characterized by four equal Chern numbers defined under twisted PBCs and a quantized shift of biorthogonal many-body polarizations under OBCs. (iv) We show generic asymmetric expansion and correlation dynamics due to the nonreciprocal hopping in the system. The AAH model has been realized with (interacting) bosonic atoms in 1D optical superlattices \cite{Roati2008,Schreiber2015,DWZhang2018,Lewenstein2007,Goldman2016,Cooper2019} and the tunable non-Hermitian gain and loss and nonreciprocal hopping have been effectively engineered for cold atoms \cite{JLi2019, WGuo2020}. Since combing these two ingredients is easily in current experiments, our results are observable. Moreover, our numerical methods can be used to explore non-Hermitian quantum many-body physics in both equilibrium and nonequilibrium cases.

The rest of this paper is organized as follows. Section \ref{sec2} introduces the 1D interacting non-Hermitian AAH model and the non-Hermitian DMRG method. In Section \ref{sec3}, we present and discuss our results, which include the stable ground states, the skin superfluid, the non-Hermitian TMIs, and the asymmetric dynamics in this system. Finally, a short conclusion is given in Section \ref{sec4}.

\section{Model and method}\label{sec2}

We start by considering a 1D optical superlattice of interacting bosons \cite{Roati2008,Schreiber2015} with nonreciprocal hoppings \cite{Hatano1996,Hatano1997}, which is described by the 1D interacting non-Hermitian AAH Hamiltonian
\begin{equation}\label{eq-ham}
\begin{aligned}
\hat{H}=&-\sum_{j}(J_r\hat{a}_{j+1}^{\dagger}\hat{a}_{j}+J_l\hat{a}_{j}^{\dagger}\hat{a}_{j+1})\\
&+V\sum_{j}\cos(2\pi\alpha j+\delta)\hat{n}_{j}+\frac{U}{2}\sum_{j}\hat{n}_{j}(\hat{n}_{j}-1),
\end{aligned}
\end{equation}
where $\hat{a}_{j}^{\dagger}$ ($\hat{a}_{j}$) is the creation (annihilation) operator of bosons at site $j$, $\hat{n}_j=\hat{a}_{j}^{\dagger}\hat{a}_{j}$ is the particle number operator, $J_r$ and $J_l$ are the nonreciprocal hopping strengths, $V$, $\alpha$ and $\delta$ denote the modulation parameters of the superlattice, and $U$ is the on-site interaction strength. We focus on the periodic modulation with $\alpha$ being a rational number and the phase $\delta\in[0,2\pi]$ acting as an effective quasimomentum in a synthetic dimension \cite{Kraus2012a,LJLang2012,Ganeshan2013}. We set $J_r=J$ and $J_l=J(1-\gamma)$, with the non-Hermiticity parameter $\gamma$ (let $\gamma\geqslant0$) and $J=1$ as the energy unit hereafter.

When $\gamma=0$, the Hamiltonian becomes Hermitian with topological insulating phases in the Mott regime \cite{SLZhu2013,XDeng2014,Kuno2017}. When $U=0$, the Hamiltonian reduces to the single-particle non-Hermitian AAH model with nontrivial topological properties \cite{QBZeng2017,Longhi2019,HJiang2019,QBZeng2019,JHou2019} (See Appendix \ref{appA}). The topological phases in this case can be characterized by nonzero Chern numbers (defined in the $k$-$\delta$ space with quasimomentum $k$ under PBCs) \cite{QBZeng2019,JHou2019}. We find a quantized shift of biorthogonal polarizations with respect to $\delta$ under OBCs as another topological invariant, which is naturally generalized to the interacting cases, as given in Eq. (\ref{P}). In the rest of this work, we explore the many-body ground states and dynamics in the general non-Hermitian interacting cases with $\gamma\neq0$ and $U\neq0$.

To numerically study 1D interacting non-Hermitian systems of size unreachable in the ED, we develop a non-Hermitian extension of the DMRG method. The DMRG method is  powerful in the numerical calculation of the ground state of a 1D strongly correlated system \cite{White1992,schollwock2011density}.
For Hermitian systems, the convergence of the ground state calculation is mainly determined by the DMRG sweeps instead of the eigensolver in each variational update. However, due to the non-orthogonality of eigenstates in non-Hermitian matrices, the convergence usually fails by merely increasing the DMRG sweeps. We solve this convergence problem by using a more accurate eigensolver, the implicitly restarted Arnoldi method~\cite{S0895479899358595}, to target to the ground state defined by the lowest real-part energy. We find that this non-Hermitian DMRG method can accurately obtain the energies and wave functions of many-body ground states. Furthermore, by generalizing the Krylov-subspace approach in Arnoldi formalism~\cite{S0036142995280572} and the time-dependent variational principle (TDVP) method ~\cite{PhysRevB.94.165116,PAECKEL2019167998}, we can simulate the non-unitary time evolution (dynamics) in this model. The ingredients and benchmarks of these non-Hermitian algorithms are presented in Appendix \ref{appB}.

In our numerical simulations, more than 200 Schmidt values in the virtual index of the MPS are kept for most conditions (400 Schmidt values are kept when the system is in the superfluid phase). The maximum local occupation is restricted to 4 and 5 bosons per site for large and small $U$s, respectively. The convergence criterion of DMRG is set to $|\Delta E_c|/E_c<10^{-8}$, where $E_c$ is the current energy and $\Delta E_c$ is the difference between previous and current energies \cite{White1992,schollwock2011density}. For the non-Hermitian TDVP method, up to 400 Schmidt values are kept during the time evolution with each time step $\tau=0.01/J$, and a maximum of 5 bosons per site is ensured. The Krylov-subspace approach to the matrix exponential applying on a vector in each local update is carried out with an exponential residual norm less than $10^{-15}$.

\section{Results and discussions}\label{sec3}

\begin{figure}[t]
\centering
\includegraphics[width=0.45\textwidth]{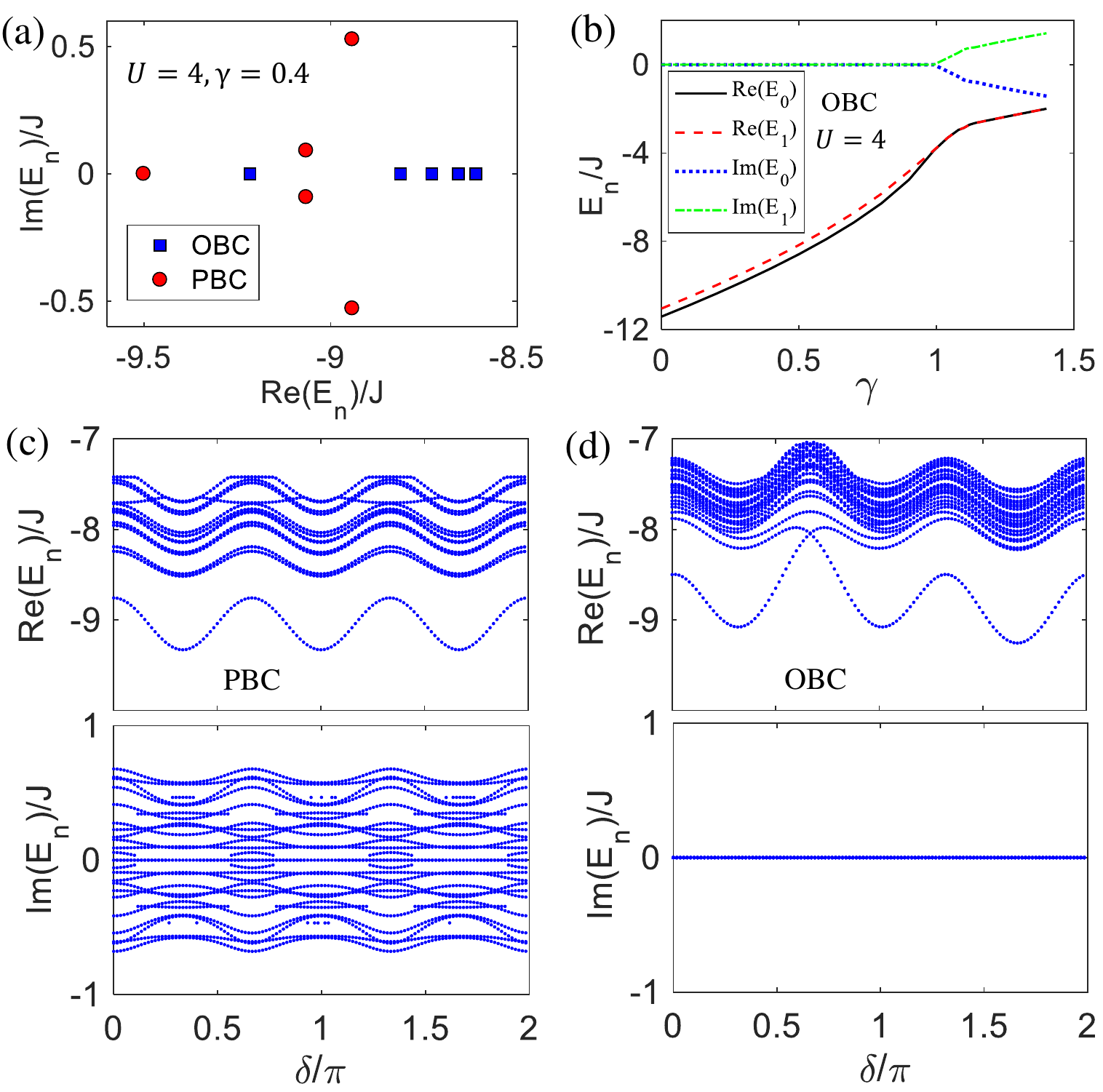}
\caption{(Color online) Energy eigenvalues of (a) five lowest states in the complex energy plane under OBCs and PBCs; and (b) two lowest states as a function of $\gamma$ under OBCs for $U=4$ and $\delta=0$. Low-energy spectra (upper for the real part and lower for the imaginary part) as a function of $\delta$ under (c) PBCs; and (d) OBCs for $U=10$. Other parmaters are $J=V=1$, $\alpha=1/3$, $L=18$, and the filling $f=1/3$.}
\label{fig1}
\end{figure}

\subsection{Stable ground states}

The complex energy spectrum $E_{n}$ and wave functions of the right (left) states $|\Psi_n^r\rangle$ ($|\Psi_n^l\rangle$) with $n=0,1,2,...$ can be obtained by solving the eigenfunction $\hat{H}|\Psi_n^r\rangle=E_n|\Psi_n^r\rangle$ ($\hat{H}^{\dag}|\Psi_n^l\rangle=E^*_n|\Psi_n^l\rangle$), which are ordered by the real-part energies $\text{Re}(E_{n})$. The ground states have the minimum value of $\text{Re}(E_{n})$ and then the excitation gap can be defined as
\begin{equation}
\Delta_{\text{ex}}=\text{Re}(E_1)-\text{Re}(E_0)
\end{equation}
for the non-degenerate ground state $|\Psi_0^r\rangle$ ($|\Psi_0^l\rangle$), as a natural extension of Hermitian systems. We numerically obtain $E_n$ for a lattice $L=18$ (ED) with $\alpha=1/3$ and the filling $f=N/L=1/3$ with the particle number $N=6$, with an example shown in Fig. \ref{fig1}(a). Under OBCs, we find that all $E_{n}$ are purely real for any $U$ when $\gamma<1$, which generally become complex when $\gamma>1$. Thus, $\gamma=1$ is the exceptional point with the parity-time symmetry breaking for the many-body ground state under OBCs, as shown in Fig. \ref{fig1}(b).

The result can be understood that there is a similarity transformation $S$ mapping the non-Hermitian Hamiltonian to a Hermitian counterpart $\hat{H}'=S\hat{H}S^{-1}$ under OBCs when $\gamma<1$, where $S$ is a diagonal matrix in the Fock space and $\hat{H}'(J',V',U')$ denotes the corresponding Hermitian interacting AAH Hamiltonian with parameters $J'=J\sqrt{1-\gamma}$, $V'=V$, and $U'=U$. The same energy spectra of $\hat{H}$ and $\hat{H}'$ are confirmed in our numerical simulations. More interestingly, we find the eigenenergies of $\hat{H}$ under PBCs are complex conjugate pais or real for any values of $\gamma$, $\delta$, and $U$; and in particular, $E_0$ is real in this case. This is guaranteed by the pseudo-Hermiticity of $\hat{H}$ under PBCs as it satisfies $\hat{I}\hat{H}\hat{I}^{-1}=\hat{H}^{\dag}$ \cite{Mostafazadeh2002,Mostafazadeh2004}, where $\hat{I}:\hat{a}_j\rightarrow\hat{a}_{L+1-j}$ is the inversion symmetry in this system. The non-degeneracy of $|\Psi_0^r\rangle$ ($|\Psi_0^l\rangle$) guarantees $E_0$ to be real. Thus, this interacting non-Hermitian system always has well-defined and stable many-body ground states with the real and smallest energy under OBCs with $\gamma<1$ or under PBCs.

We also calculate the low-energy spectra of the system Hamiltonian as a function of the modulation phase $\delta$ under both PBCs and OBCs, as shown in Figs. \ref{fig1}(c) and \ref{fig1}(d). We numerically confirm that for all $\delta$, the energy of the ground state [with smallest $\text{Re}(E_{n})$] is always real under PBCs and the whole energy spectrum is real under OBCs. In addition, by taking $\delta$ as an effective quasimomentum (periodic along this artificial dimension) and finite lattice site under OBCs in real space, one can find that the ground state and the excited states cross near $\delta=2\pi/3$ in the real bulk gap, where the gapless excitations emerge. The edge localization and the topological nature of these gapless excitations in the Mott insulating phase will be studied in Sec. \ref{Sec-TMI}.

\subsection{Skin superfluid under OBCs}

\begin{figure}[t]
\centering
\includegraphics[width=0.45\textwidth]{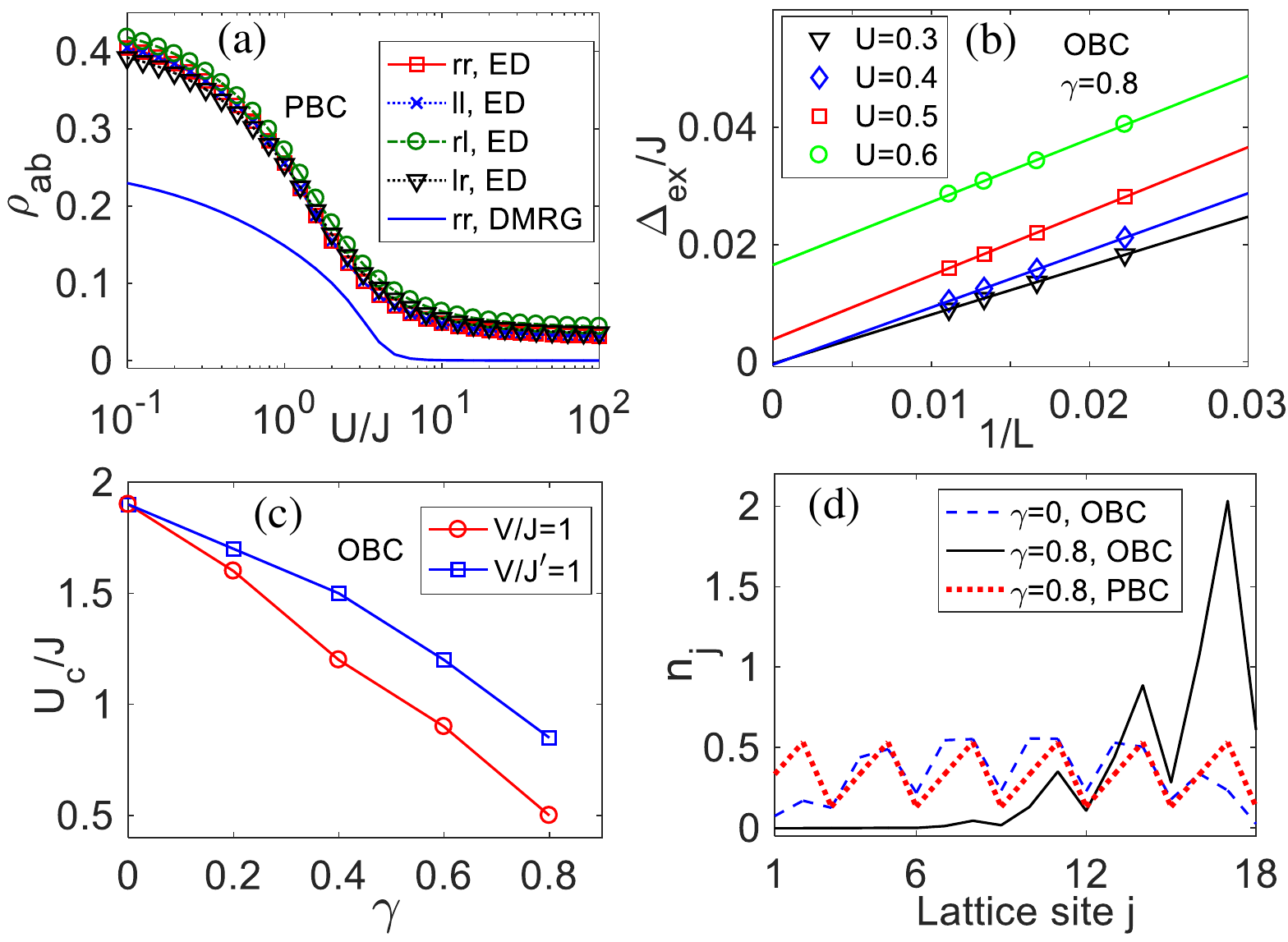}
\caption{(Color online) (a) One-particle density matrix $\rho_{ab}$ as a function of $U$ for $L=18$ (ED) and $L=90$ (DMRG) under PBCs, respectively. (b) Finite-size scaling of $\Delta_{\text{ex}}$ for various $U$ (the labels) and $\gamma=0.8$ under OBCs. (c) Critical point $U_c/J$ as a function of $\gamma$ for $V/J=1$ and $V/J'=1$, respectively. (d) Density distribution $n_j$ of the superfluid ground state $|\Psi_0^r\rangle$ for $U=0.4$. Other parameters are $J=V=1$, $\alpha=1/3$, $\delta=0$ and the filling $f=1/3$.}
\label{fig2}
\end{figure}

The many-body ground state of the interacting bosons is in the superfluid or Mott insulating phase with a critical interacting strength $U_c$. We now consider the superfluid state, which can be characterized by the one-particle density matrix. For generic non-Hermitian systems, four components of the one-particle density matrix can be defined according to whether left and right eigenstates are assigned in the expectation value \cite{SMu2019,Yamamoto2019}, which are given by $\rho_{ab}=\langle \Psi_{0}^{a} |\hat{a}_1^{\dagger} \hat{a}_{L/2+1}|\Psi_{0}^b\rangle$ for right and left ground states under PBCs, with $a,b=l/r$. Here $\rho_{rr}$ ($\rho_{ll}$) characterizes only the ground state $|\Psi_{0}^r\rangle$ ($|\Psi_{0}^l\rangle$) governed by the Hamiltonian $H$ ($\hat{H}^{\dag}$), and the biorthogonal component $\rho_{rl}$ ($\rho_{lr}$) can give the probability-conserving expectation value. As we are interested in the system governed by $H$, we can focus on the ground state $|\Psi_{0}^r\rangle$. Under PBCs, $\rho_{ab}$ can capture the momentum distribution of the interacting bosons and is associated with an off-diagonal quasi-long-range order if $\rho_{ab}$ remains finite at large $L$. The typical results of $\rho_{ab}$ from the ED ($L=18$) and the DMRG ($L=90$) as a function of $U$ are shown in Fig. \ref{fig2}(a). The four components $\rho_{ab}$ are real (here the slight differences among $\rho_{ab}$ are due to the modulation potential $V$) and indicate the superfluid (Mott insulating) phase for small (large) $U/J$, similar as those in Hermitian Bose-Hubbard models \cite{Kuhner2000}. In the case of OBCs, the four components $\rho_{ab}$ are dramatically different because the asymmetric hopping can induce the
accumulation of the right (left) state $|\Psi_{0}^r\rangle$ ($|\Psi_{0}^l\rangle$) to the right (left) boundary (see Fig. \ref{fig2}(b) for such a non-Hermitian skin effect \cite{SYao2018}). To better characterize the two phases of the model Hamiltonian $H$ under OBCs, we can calculate the excitation gap for the right states.

The excitation gap can be used to determine the critical point $U_c/J$ (at zero temperature and in the thermodynamic limit) between the gapless superfluid phase and the gapped Mott insulting phase. For realistic systems under OBCs, the excitation gap $\Delta_{\text{ex}}$ for the non-Hermitian Hamiltonian $\hat{H}(J,U,V)$ can be obtained from that for the Hermitian counterpart $\hat{H'}(J',U,V)$ after the similarity transformation when $\gamma<1$. Figure \ref{fig2}(b) shows the finite-size scaling of $\Delta_{\text{ex}}$ for various $U$ and $\gamma=0.8$, which gives $U_{c}/J\sim0.5$ through extrapolation to the $L\rightarrow \infty$ limit. Using this procedure, we numerically obtain $U_{c}/J$ as a function of the non-Hermiticity $\gamma$ for fixed $V/J=1$ or $V/J'=1$, as shown in Fig. \ref{fig2}(c). One can find that $U_c/J$ is decreased when increasing $\gamma$ and $U_c/J\propto\sqrt{1-\gamma}$ for the case with fixed $V/J'$, which can be understood from the reduced effective reciprocal hopping $J'=J\sqrt{1-\gamma}$ under OBCs.

In Fig. \ref{fig2}(d), we show the superfluid density distribution $n_j=\langle \Psi_{0}^{r} |\hat{n}_j|\Psi_{0}^r\rangle$ defined for the right ground state for different conditions. Under PBCs, we find that the superfluid density distribution is periodically modulated by the superlattice potential $V$, which is independent on $\gamma$ and will become nearly uniform when $V/J\ll1$. Under OBCs, in contrast, the superfluid tend to localized at the right side of the lattice by increasing $\gamma$ ($0\leqslant\gamma<1$) due to the asymmetric hopping $J_r>J_l$. This can be understood as a many-body generalization of the non-Hermitian skin effect \cite{SYao2018}, and thus such a superfluid state under OBC is dubbed skin superfluid. Notably, the skin superfluid under OBCs is more significant for larger $\gamma$ and smaller $U/J$, and the potential $V$ just adds the overall modulation in the density distribution.

\subsection{Non-Hermitian TMIs} \label{Sec-TMI}

\begin{figure}[t]
\centering
\includegraphics[width=0.45\textwidth]{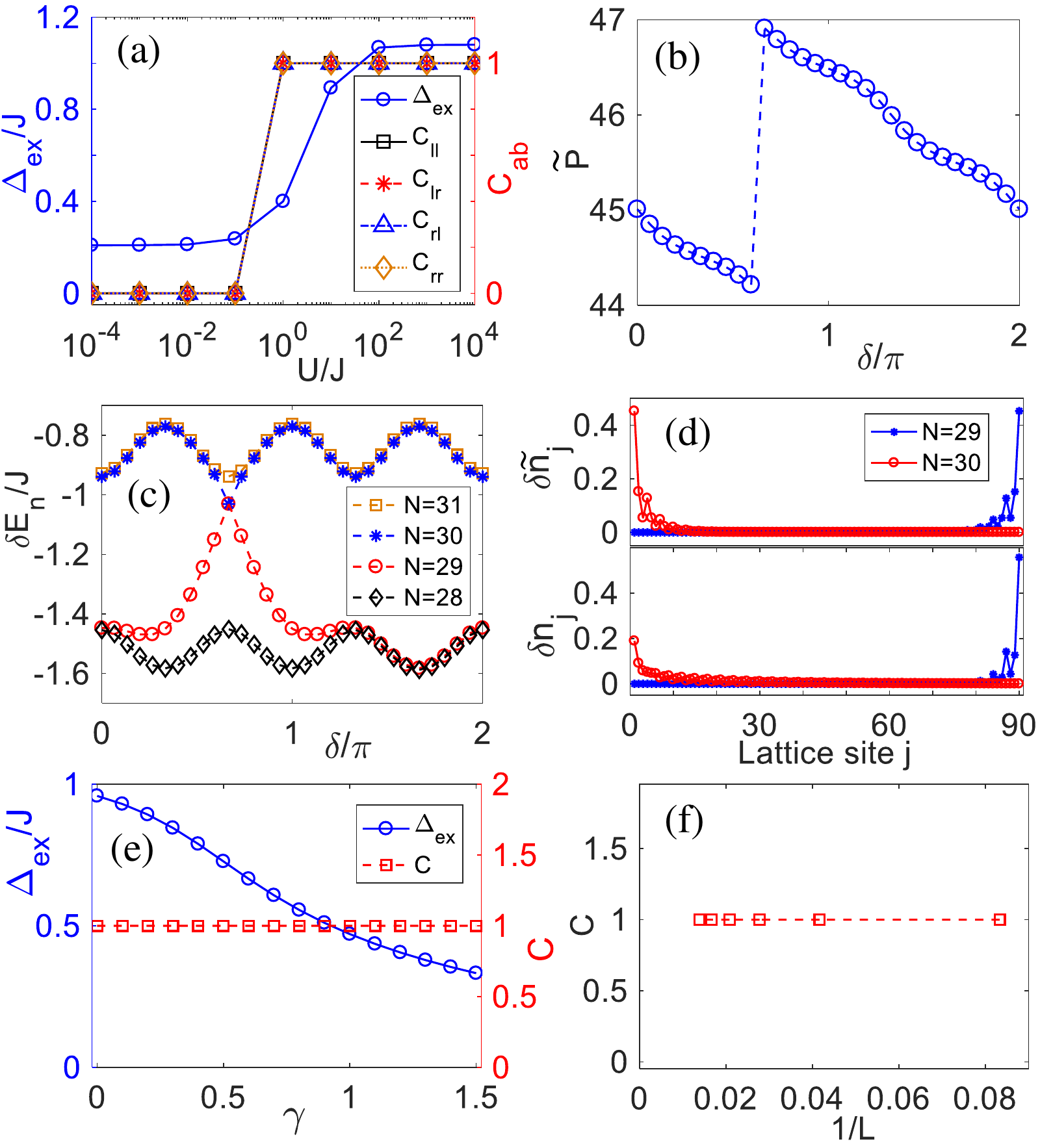}
\caption{(Color online) Excitation gap $\Delta_{\text{ex}}$ and four Chern numbers $C_{ab}=C$ as a function of (a) $U$ and (e) $\gamma$, for $L=12$ and $N=4$ obtained from the ED under PBCs. (b) Biorthogonal polarization $\tilde{P}$ as a function of $\delta$; (c) Four middle branches of the quasiparticle energy spectrum $\delta E_N$; and (d) Biorthogonal and right-state density distributions $\tilde{n}_j$ and $n_j$ of the two in-gap modes with $\delta=2\pi/3$, obtained from the DMRG for $L=90$ under OBCs. (f) Finite-size scaling of $C$ from the DMRG. Other parameters are $J=V=1$, $U=10$, $\alpha=1/3$, and $\gamma=0.2$.}
\label{fig3}
\end{figure}

We proceed to investigate the topological properties of the Mott insulating phase. Figure \ref{fig3}(a) depicts that the excitation gap $\Delta_{\text{ex}}$ increases as a function of $U$ and saturates at a finite value for large $U$. Note that the nonzero value of $\Delta_{\text{ex}}$ for the superfluid phase in the small $U$ limit is actually  the finite-size gap in the ED simulation. In the presence of a finite gap, we can use the Chern number to characterize the topology of the Mott insulators. For our non-Hermitian system, under the twisted PBCs with a twisted phase $\theta$ \cite{QNiu1985}, we can define four Chern numbers in the $\theta$-$\delta$ space as
\begin{equation}
C_{ab}=\frac{1}{2\pi}\int_{0}^{2\pi}d\theta \int_{0}^{2\pi}d\delta F_{ab}(\theta, \delta),
\end{equation}
where $F_{ab}(\theta,\delta)=i\langle\partial_{\theta}\Psi_0^{a}|\partial_{\delta}\Psi_0^{b}\rangle$ is the four Berry curvatures related to the left and right many-body ground states. We calculate $C_{ab}$ by evaluating $F_{ab}$ using a discrete manifold method \cite{Fukui2005}, and find that four Chern numbers are equal $C_{rr}=C_{ll}=C_{rl}=C_{lr}=C$ as a function of $U$ as shown in Fig. \ref{fig3}(a). This implies that the Chern number $C$ is invariant to different choices of right and left many-body ground states \cite{HShen2018}. In the deep Mott insulator regime, such as $U=10$, one has $C=1$ indicating a TMI. Note that the Chern number is actually not well-defined for small $U$ although $C$ remains quantized in Fig. \ref{fig3}(a) due to the finite-size effect.

\begin{figure*}[t]
\centering
\includegraphics[width=0.8\textwidth]{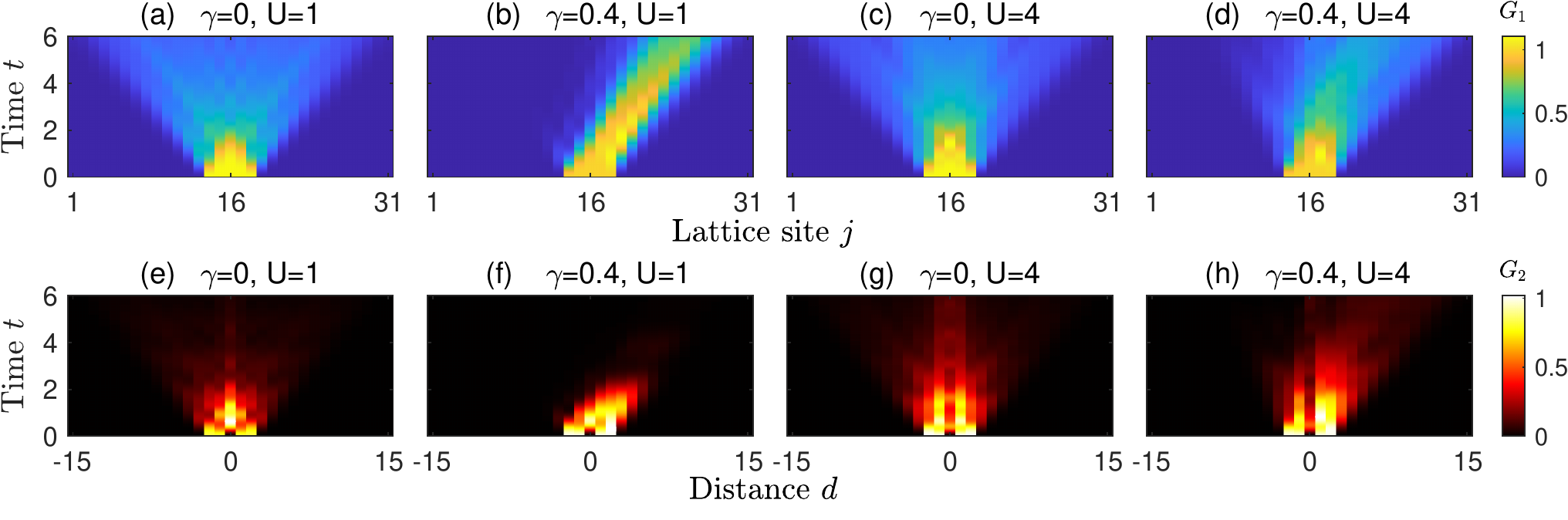}
\caption{(Color online) Dynamics of one-particle density $G_1$ and two-particle correlation $G_2$ for the initial state with $N=5$ bosons localized one per site in the center of a $L=31$ lattice. (a-d) $G_1$ with zero and non-zero asymmetric hopping $\gamma$, and moderate and strong interaction strength $U$. (e-h) $G_2$ for the same parameters with respect to the upper panel (a-d).}
%Other parameters are $J=1$ and $V=0$.}
\label{fig4}
\end{figure*}

We find another topological invariant to characterize the TMI under OBCs, which is related to the biorthogonal polarization $\tilde{P}$ \cite{Kunst2018,Lee2019a}. For the many-body ground state under OBCs, the expression of $\tilde{P}$ for $L$ lattice sites and $N$ bosons is given by
\begin{equation} \label{P}
\tilde{P}(\delta)=\frac{1}{N}\sum_{j=1}^{L}j\langle\Psi_0^l(\delta)|\hat{n}_j|\Psi_0^r(\delta)\rangle,
\end{equation}
which is a function of the periodical modulation phase $\delta$. Figure \ref{fig3}(b) depicts the DMRG results of $\tilde{P}(\delta)$ for $L=3N=90$, $\gamma=0.2$ and $U=10$. Here $\tilde{P}$ exhibits a jump of nearly one unit cell (three sites) by varying $\delta$ from $0$ to $2\pi$, corresponding to $C=1$. So a quantized shift (when $L\rightarrow\infty$) of the biorthogonal many-body polarization under OBCs is also a topological invariant for our interacting non-Hermitian model.

The quasiparticle energy spectrum under OBCs can be obtained from the DMRG, which is real and given by $\delta E_N=E_{0,N+1}-E_{0,N}$. Here $E_{0,N}$ is the energy of the ground state with $N$ bosons denoted by $|\Psi_{0,N}^r\rangle$ and is real under OBCs ($\gamma<1$). %Note that $\delta E_N$ denotes the extra energy needed to add a boson to the system with $N$ bosons.
Figure \ref{fig3}(c) shows $\delta E_N$ as a function of $\delta$ near the filling $f=1/3$ for $L=90$. There are two branches of in-gap edge modes that are degenerate at $\delta=2\pi/3$ and connect the lower and the upper bulk spectra when $\delta$ varies from $0$ to $2\pi$, corresponding to the TMI with $C=1$. The right-state and biorthogonal density distributions of quasiparticles are respectively given by $\delta n_j =\langle\Psi_{0,N+1}^r|\hat{n}_j|\Psi_{0,N+1}^r\rangle-\langle\Psi_{0,N}^r|\hat{n}_j|\Psi_{0,N}^r\rangle$ and $\delta \tilde{n}_j =\langle\Psi_{0,N+1}^l|\hat{n}_j|\Psi_{0,N+1}^r\rangle-\langle\Psi_{0,N}^l|\hat{n}_j|\Psi_{0,N}^r\rangle$. The results of $\delta n_j$ and $\delta \tilde{n}_j$ for the two degenerate edge modes at $\delta=2\pi/3$ are shown in Fig. \ref{fig3}(d). Due to the asymmetric hopping for the right eigenstates, $\delta n_j$ for the two edge modes exhibit asymmetric distributions. The asymmetric hopping is cancelled under the biorthogonal eigenstates and then $\delta \tilde{n}_j$ for the two edge modes remain symmetric distributions.

We further study the non-Hermitian effect on the TMI. Figure \ref{fig3}(e) shows that the obtained Chern number is independent on the non-Hermiticity $\gamma$. %from $\gamma=0$.
Remarkably, it preserves even when $\gamma>1$ as the ground state still has a finite gap in the complex energy plane under PBCs. Thus, although the many-body energy spectrum is generally complex, the non-Hermitian TMI in this system can still be topologically connected to the Hermitian TMI with $\gamma=0$. This can be understood that 1D interacting non-Hermitian systems share the same topological classification as that of the Hermitian systems \cite{WXi2019}. We further confirm that the Chern number and the stable ground states with real energies preserve in the large $L$ limit from the DMRG, as shown in Fig. \ref{fig3}(f).

\subsection{Asymmetric dynamics}
Based on the TDVP method, we can investigate the dynamics in 1D non-Hermitian many-body systems (see Appendix \ref{appB}). We find that the nonreciprocal hopping can induce asymmetric expansion and correlation dynamics in this system. We consider the time-evolution of the one-particle density $G_1(j,t)=\braket{\Psi^r(t)|\hat{a}^\dagger_j\hat{a}_j|\Psi^r(t)}$ and the two-particle correlation function $G_2(q,p,t)=\braket{\Psi^r(t)|\hat{a}^\dagger_q\hat{a}^\dagger_p\hat{a}_p\hat{a}_q|\Psi^r(t)}$~\cite{PhysRevLett.105.163905} to show the generic asymmetric dynamics, which are defined for the right states $\ket{\Psi^r(t)}$. Note that $G_1$ and $G_2$ (related to the density-density correlation) are both measurable for ultracold bosons in optical latices. Typically, we consider the initial state $\ket{\Psi^r(0)}$ as a product state in the Fock space with $N=5$ bosons localized at the center five sites of a lattice with $L=31$. The system is then driven by the non-Hermitian Hamiltonian $\hat{H}$ in Eq.~\ref{eq-ham}.

The time-evolved state $\ket{\Psi^r(t)}=\mathrm{e}^{-i\hat{H}t}\ket{\Psi^r(0)}$ is calculated with the TDVP method with $J=1$ and $V=0$. %is assumed.
We fix the lattice site $p=16$ at the center of the lattice and then rewritten the two-particle correlation $G_2(d,t)=\braket{\Psi^r(t)|\hat{a}^\dagger_{p+d}\hat{a}^\dagger_p\hat{a}_p\hat{a}_{p+d}|\Psi^r(t)}$ by introducing the distance $d=q-p$. The expansion dynamics of $G_1$ and the correlation dynamics of $G_2$ for zero and non-zero asymmetric hoppings $\gamma$ and two interaction strengths $U$ are shown in upper and lower panels of Fig.~\ref{fig4}, respectively. In the absence of asymmetric hopping $\gamma$ [Figs.~\ref{fig4} (a,c,e,g)], the dynamics of both $G_1$ and $G_2$ are symmetric with respect to the center of the lattice. For the non-Hermitian cases, the one-particle density $G_1$ prefers to propagate to the right-hand side of the lattice [Figs.~\ref{fig4} (b,d)] as the hopping $J_r>J_l$. This preference can be observed in the superfluid and Mott insulating phases for the two calculated interacting strengths. Similar asymmetric dynamics occur in the two-particle correlation $G_2$ [Figs.~\ref{fig4} (f,h)]. At the same time $t$, $G_2$ for $d>0$ enjoys larger value than that for $d<0$. It means that it is more likely to detect particles simultaneously at the center and right-hand side of the lattice than that at the left-hand side. Notably, the asymmetric expansion and correlation dynamics are absent under the biorthogonal eigenstates since the asymmetric hopping is cancelled in this basis.

\section{Conclusions}\label{sec4}

In summary, we have explored the stable many-body ground states and quantum dynamics in the 1D interacting non-Hermitian AAH model. We have revealed the nonreciprocal-hopping-induced skin superfluid under OBCs and the TMIs independence of non-Hermiticity. The TMIs are characterized by four equal Chern numbers and a quantized shift of biorthogonal many-body polarizations. We have also shown generic asymmetric expansion and correlation dynamics due to the nonreciprocal hopping in this system. The AAH model can be realized with interacting ultracold bosons in 1D optical suerlattices \cite{Roati2008,Schreiber2015} and the effective nonreciprocal hopping can be engineered by using an atomic one-body loss \cite{ZGong2018,JLi2019,WGuo2020}. Thus, our predicted results could be observed in cold atom experiments. Furthermore, our numerical methods are applicable to explore non-Hermitian quantum many-body physics in both equilibrium and nonequilibrium cases.

\textit{Note added:} After our submission, we noticed two complimentary works, which focused on the non-Hermitian TMIs of interacting fermions and bosons \cite{TLiu2020,ZXu2020}.% where the superfluid state, the dynamics, and the non-Hermitian DMRG were not considered.

\begin{acknowledgments}
This work was supported by the NSFC (Grants No. 11704367, No. 11904109, No. 91636218, No. U1830111, and No. U1801661), the NKRDP of China (Grant No. 2016YFA0301800), the Key-Area Research and Development Program of Guangdong Province (Grant No. 2019B030330001), and the Key Program of Science and Technology of Guangzhou (Grant No. 201804020055).

D.W.Z., Y.L.C., and G.Q.Z contributed equally to this work.

\begin{appendix}

\section{Single particle physics}\label{appA}

\begin{figure*}[t]
	\centering
	\includegraphics[width=0.7\textwidth]{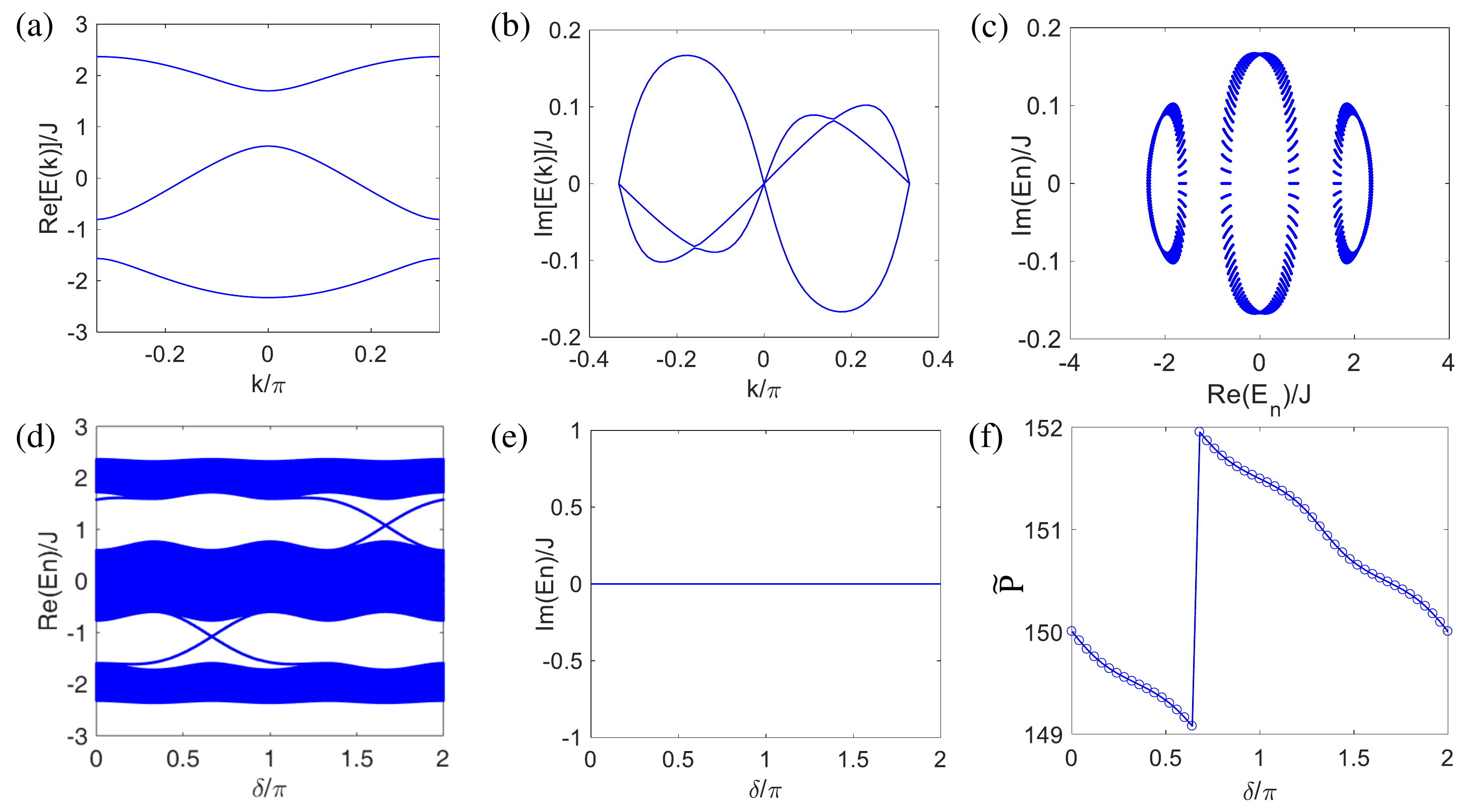}
	\caption{(Color online) (a)-(e) Energy spectra of the single-particle non-Hermitian AAH model with $V=1$, $\gamma=0.2$ and $\alpha=1/3$. (a) The real and (b) imaginary part of the energy bands under PBCs. (c) The energy spectrum in the complex plane under PBCs. (d) The real and (e) imaginary part of the energy spectrum under OBCs as a function of $\delta$ with $L=300$. (f) The biorthogonal polarization $\tilde{P}$ as a function of $\delta$ under OBCs.}
	\label{figS1}
\end{figure*}

In the non-interacting limit, the model Hamiltonian in Eq. (\ref{eq-ham}) becomes
\begin{equation}\label{eq-spham}
\hat{H}_0=-\sum_{j}(J_r\hat{a}_{j+1}^{\dagger}\hat{a}_{j}+J_l\hat{a}_{j}^{\dagger}\hat{a}_{j+1})+V\sum_{j}\cos(2\pi\alpha j+\delta)\hat{n}_{j},
\end{equation}
which is a non-Hermitian AAH model. The $n$-th eigenstate of the Hamiltonian is given by $|\Psi_n^r \rangle = \sum_{j}u_{j,n}c_j^{\dagger}|0\rangle$ with the eigen equation $\hat{H}_0|\Psi_n^{r}\rangle=E_n|\Psi_n^{r}\rangle$, where $u_{j,n}$ is the wave function at the $j$-th site with the eigenenergy $E_n$. We can obtain the generalized Harper equation
\begin{equation}\label{Harper1}
-(J_ru_{j+1,n}+J_lu_{j-1,n})+V_ju_{j,n}=E_nu_{j,n}.
\end{equation}
In the commensurate case of rational $\alpha$ and periodic $V_j$ with a period $Q$, one can suppose $\psi_j(k)~(j=1,\cdots,Q)$ as the wave function in the momentum space and take $u_j=e^{ikj}\psi_{j}(k)$ for $k\in[-\pi/Q,\pi/Q]$, then the Harper equation in Eq. (\ref{Harper1}) becomes
\begin{equation}\label{Harper2}
	-(J_re^{ik}\psi_{j+1}+J_le^{-ik}\psi_{j-1})+V_j\psi_{j}=E(k)\psi_{j}.
\end{equation}
Under PBCs, we can obtian the eigenenergies and eigenstates by diagonalizing the following Hamiltonian matrix
$$
	\begin{bmatrix}
	V_1        &J_le^{-ik} &0          &\cdots    &J_re^{ik}\\
	J_re^{ik}  &V_2        &J_le^{-ik} &\cdots    &0  \\
    \vdots     &\ddots     &\ddots     &\ddots    &\vdots \\
    0          &\cdots     &J_re^{ik}  &V_{Q-1}    &J_le^{-ik}\\
	J_le^{-ik} &\cdots      &0         &J_re^{ik}  &V_Q
	\end{bmatrix}.
$$

In Fig. \ref{figS1}, we plot the energy spectra in the complex plane for varying $\delta$ from $0$ to $2\pi$ with $\alpha=1/3$ under PBCs. We show that the real part of the energy spectrum consists of three bands in Fig. \ref{figS1}(a). In addition, there are non-zero imaginary part of the energy in the non-Hermitian AAH model, as shown in Fig. \ref{figS1}(b). Since the energy spectrum is separated in the complex plane, the Chern number is still well-defined for this non-Hermitian system \cite{HShen2018}, which is defined in the $k$-$\delta$ space in this case. Notably, we find that the Chern number is quantized to be exactly 1 for the lowest band. For OBCs, we calculate the eigenenergy and the biorthogonal polarization of the system. The biorthogonal polarization is given by \cite{Kunst2018}
$\tilde{P}(\delta)=\frac{1}{N}\sum_{j=1}^{L}j{|\Psi_j(\delta)|^2}$,
where the density distribution $|\Psi_j(\delta)|^2 $ of the occupied states at site $j$ is calculated in the biorthogonal basis. As shown in Figs. \ref{figS1}(d) and \ref{figS1}(e), the non-Hermitian system has real energy spectrum under OBCs and the edge modes appear in the gapped regime. We also see the biorthogonal polarization $\tilde{P}$ exhibits a change of nearly one unit cell when $\delta$ varies from $0$ to $2\pi$ in Fig. \ref{figS1}(f). Similar as that in the Hermitian case, the change of the biorthogonal polarization is proportional to the Chern number.

\section{Numerical methods and benchmarks} \label{appB}

The DMRG is one of the most powerful numerical methods for 1D strongly correlated systems ~\cite{White1992,schollwock2011density}. %However, its flexibility for non-Hermitian systems remain unclear and the non-orthogonal eigenstates require a more accurate eigensolver for each variational update. By adapting the implicitly restarted Arnoldi method~\cite{S0895479899358595}, we report that the non-Hermitian DMRG method can accurately obtain the energies and wave functions of the many-body ground state, which is sufficient to reveal the quantum phases and topological properties in our model. Moreover, we employ the time-dependent variational principle (TDVP) method ~\cite{PhysRevB.94.165116,PAECKEL2019167998} to investigate the dynamics of the system by numerical integration of the time-dependent Schr\"{o}dinger equation, and the Krylov-subspace approach in Arnoldi formalism~\cite{S0036142995280572} is adopted for the non-unitary time evolution. %$\ket{\Psi_R(t+\delta t)}=\mathrm{e}^{-i\hat{H}\delta t}\ket{\Psi_R(t)}$ with the time step $\delta t$.
%%Based on the matrix-product state (MPS) representation, the TDVP method can reach large systems beyond the ED method using sparse matrix exponentials.
Here we discuss the technical ingredients of the non-Hermitian DMRG and TDVP methods and benchmark the flexibility of these numerical methods for the system we investigated, which has real or complex ground state energy at different parameters. Let us begin with the matrix product state (MPS) representation of a general 1D quantum state $\ket{\Psi}$:
\begin{equation}
\begin{aligned}
\ket{\Psi}=&\sum_{j_1,\cdots,j_L}\sum_{a_0,a_1,\cdots,a_{L}}\mathrm{Tr} (M^{j_1}_{a_0,a_1}M^{j_2}_{a_1,a_2}\cdots \\ &M^{j_{L-1}}_{a_{L-2},a_{L-1}}M^{j_L}_{a_{L-1},a_L})\ket{j_1,j_2,\cdots,j_{L-1},j_{L}}.
\end{aligned}
\end{equation}
Each $M^{j_n}_{a_{n-1},a_n}$ is a rank three tensor with $j_n$ standing for the index of a local state, and $a_n$ is a virtual index which connects adjacent sites (we will omit virtual indices for the sake of simplicity). Under the OBCs, $a_0$ and $a_L$ are dummy indices and the trace operation is not needed. There exists a gauge freedom and one can bring the MPS into canonical form:
\begin{equation}
\ket{\Psi}=\sum_{\alpha=1}^{\chi_{n}} \Lambda_{\alpha \alpha}^{[n]}\ket{l_\alpha^{[n]}}\ket{r_\alpha^{[n]}},
\end{equation}
where $n$ is the canonical center, $\Lambda^{[n]}$ is a diagonal matrix containing the Schmidt values of the bipartition $\{l,r\}$, $\chi_n$ is the total number of Schmidt values kept, $\ket{l_\alpha^{[n]}}=\sum_{j_1,\cdots,j_{n}}(A^{j_1}\cdots A^{j_n})_\alpha\ket{j_1,\cdots,j_n}$ and $\ket{r_\alpha^{[n]}}=\sum_{j_{n+1},\cdots,j_{L}}(B^{j_{n+1}}\cdots B^{j_L})_\beta\ket{j_{n+1},\cdots,j_L}$ are the orthonormal Schmidt states of the left and right parts respectively, and $A^{j_i}$ ($B^{j_i}$) is the left (right) canonical form of the $M^{j_i}$ tensor at site $i$. The two-site DMRG method converges to the ground state $\ket{\Psi_0}$ by variationally optimizing two neighboring MPS tensors $M^{j_n,j_{n+1}}=A^{j_n}\Lambda^{[n]}B^{j_{n+1}}$ at once and minimize the energy $\braket{\Psi_0|\hat{H}|\Psi_0}$. The simply roadmap of the two-site algorithm is as follows~\cite{schollwock2011density}:
\begin{enumerate}
\item Prepare the MPS wavefunction $\ket{\Psi_0}$ with canonical center at site $n$:
\begin{equation*}
\ket{\Psi_0}=\sum_{\alpha,j_n,j_{n+1},\beta}M^{j_n,j_{n+1}}\ket{l_\alpha^{[n-1]}}\ket{j_n}\ket{j_{n+1}}\ket{r_\beta^{[n+1]}}.
\end{equation*}
In the variational update, tensors belong to $\ket{l^{[n-1]}}$ and $\ket{r^{[n+1]}}$ are fixed and $M^{j_n,j_{n+1}}=A^{j_n}\Lambda^{[n]}B^{j_{n+1}}$ should be improved.

\item Generate the effective Hamiltonian $\hat{H}_{\mathrm{eff}}$ under the projected basis $\ket{l_\alpha^{[n-1]}j_nj_{n+1}r_\beta^{[n+1]}}$.

\item Numerically find the lowest lying eigenvector $\tilde{M}^{j_n,j_{n+1}}$ of the effective Hamiltonian $\hat{H}_{\mathrm{eff}}$.

\item (Sweep from left) Update tensor $A^{j_n}=\tilde{A}^{j_n}$ on sites $n$ from the singular value decomposition (SVD) of $\tilde{M}^{j_n,j_{n+1}}=\tilde{A}^{j_n}\tilde{\Lambda}^{[n]}\tilde{B}^{j_{n+1}}$, prepare tensor $M^{j_{n+1},j_{n+2}}=\tilde{\Lambda}^{[n]}\tilde{B}^{j_{n+1}}B^{j_{n+2}}$ for the next pair of sites.

(Sweep from right) Update tensor $B^{j_{n+1}}=\tilde{B}^{j_{n+1}}$ on sites $n+1$ from the singular value decomposition of $\tilde{M}^{j_n,j_{n+1}}=\tilde{A}^{j_n}\tilde{\Lambda}^{[n]}\tilde{B}^{j_{n+1}}$, prepare tensor $M^{j_{n-1},j_{n}}=A^{j_{n-1}}\tilde{A}^{j_{n}}\tilde{\Lambda}^{[n]}$ for the next pair of sites.

\item Return to step 2 and sweep through the whole system till convergence.
\end{enumerate}

\begin{figure*}[t]
\centering
\includegraphics[width=0.7\textwidth]{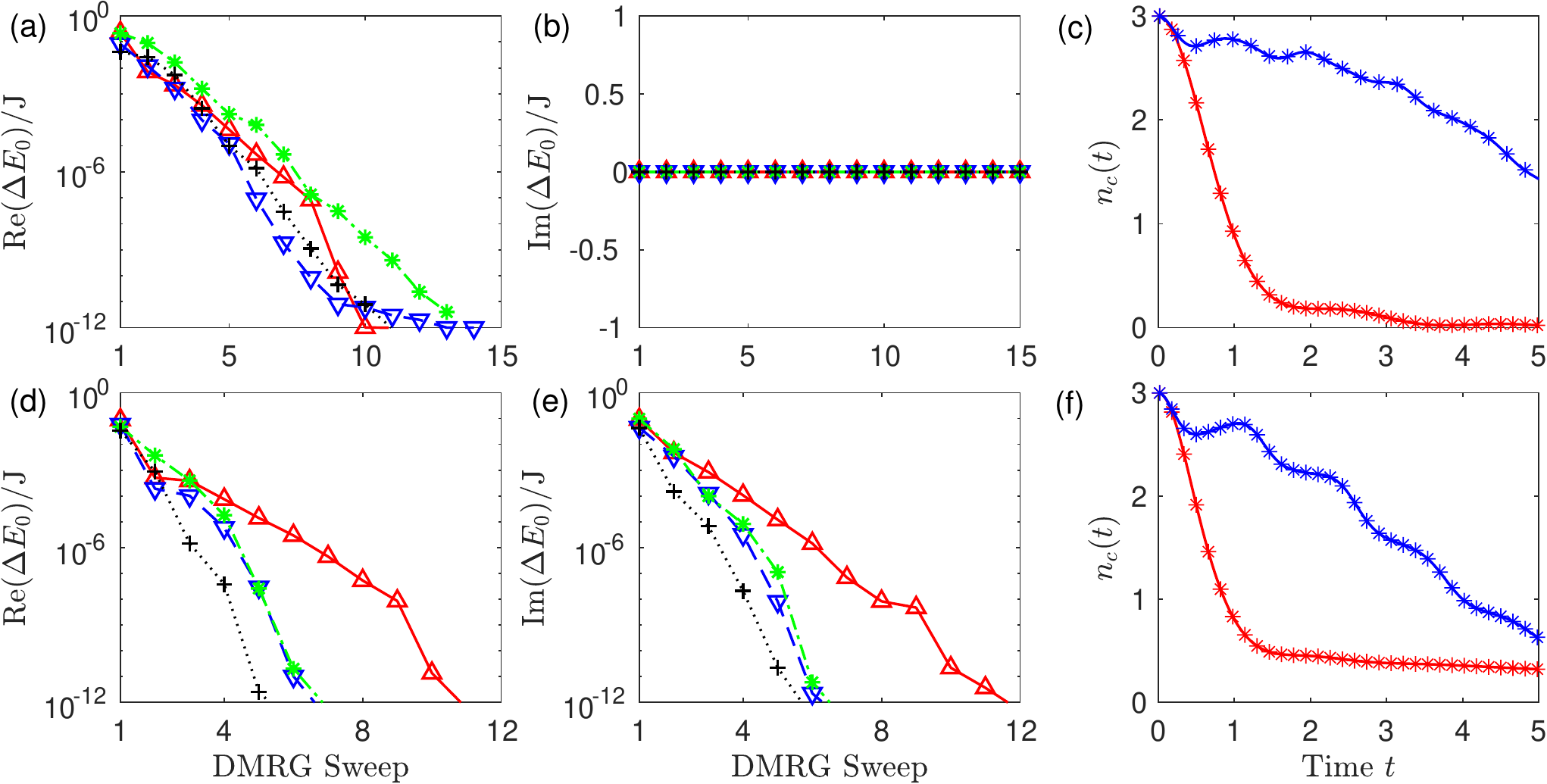}
\caption{(Color online) (a)-(b) Real and imaginary parts of the ground state energy error $\Delta E_0$ as a function of the DMRG sweep, parameters are chosen as: $\gamma=0.2$, $U=10$, $L=90$ (red solid lines); $\gamma=0.2$, $U=4$, $L=45$ (blue dashed lines); $\gamma=0.4$, $U=4$, $L=45$ (green dash-dotted lines); $\gamma=0.8$, $U=10$, $L=30$ (black dotted lines); other parameters are $J=1$, $V=1$, $\alpha=1/3$, $\delta=2\pi/3$ and the filling $f=2/3$. (c) The expected occupation on the central site $c=11$ for a $L=21$ lattice as a function of time $t$ obtained by the ED (asterisk points) and TDVP (solid curves) methods, the initial state is $N=3$ bosons locate in the central site. Parameters are chosen as: $\gamma=0.4$, $U=1$ (red curve); $\gamma=0.4$, $U=4$ (blue curve); other parameters are $J=1$, $V=0$. (d)-(f) The same as (a)-(c) with a different nonreciprocal hopping strength $J_L=e^{i\gamma}J$ where the system no longer has real ground state energy.}
\label{figS2}
\end{figure*}

The non-orthogonality of a general matrix requires a more accurate eigensolver in step 3 of the non-Hermitian DMRG method and we implement a complex general version of the implicitly restarted Arnoldi (IRA) method similar to the widely used ARPACK package~\cite{lehoucq1998arpack}. In practice, the IRA method only requires the action of $\hat{H}_{\mathrm{eff}}$ on a vector, and steps 2 and 3 are combined to avoid an explicit computation and storage of the effective Hamiltonian $\hat{H}_{\mathrm{eff}}$ which will significantly reduce the performance of the algorithm. The IRA method is set to target the eigenvalue with smallest real part in the spectrum, and typically $3\sim5$ Arnoldi basis vectors and a strict energy toleration are used during numerical simulations.

In Figs.~\ref{figS2} (a) and (b), we plot the real and imaginary parts of the targeted ground state energy error $\Delta E_0$ for the 1D interacting non-Hermitian AAH Hamiltonian during the non-Hermitian DMRG sweeps, respectively. The energy error is defined as the difference between the previous energy $E^\prime_0$ and current energy $E_0$ reported by the DMRG algorithm: $\Delta E_0=E^\prime_0-E_0$. For strong interacting strength $U=10$, the non-Hermitian DMRG method can quickly converge in a $L=90$ lattice, and for moderate interacting strength $U=4$, it also has good convergence in a lattice of $L=45$. In this benchmark, a moderate $L$ can convergent even for a very large non-Hermitian parameter $\gamma=0.8$. The reachable lattice size for small $U$ is restricted because the energy gap between the ground state and excited state becomes difficult to distinguish in the variational optimization. In all cases benchmarked, asymmetric hopping strengths $\gamma=0.2$, $\gamma=0.4$, and $\gamma=0.8$ show zero imaginary parts during the DMRG sweeps. We also benchmark systems with non-zero imaginary ground state energy and present results in Figs.~\ref{figS2} (d) and (e), where the nonreciprocal hopping strengths $J_r=J$ and $J_l=e^{i\gamma}J$. Parameters are chosen the same as Figs.~\ref{figS2} (a,b) and both real and imaginary parts convergent to $\Delta E_0=10^{-12}$ for all cases. The PBCs for asymmetric hopping systems are also benchmarked with the same parameters as shown  Fig.~\ref{figS2}(a,b). The zero imaginary parts are ensured during all sweeps, and the real parts of the ground state energy error $\Delta E_0$ is about two orders of magnitude larger than the OBC counterpart, which is due to the fact that the systems under PBCs usually take more sweeps for convergence in DMRG simulations.

%Based on the obtained eigenenergies and wave functions of the many-body ground state from the DMRG, we further confirm that the Chern numbers and the stable ground states with real energies preserve in the large $L$ limit. The results of an example is shown in Fig. \ref{figS3}, where the parameters are chosen the same as Figs. 3(e,f) from the ED.

%\begin{figure}[t]
%\centering
%\includegraphics[width=0.6\textwidth]{FigS3.pdf}
%\caption{(Color online) (a) Chern number $C$ and (b) ground-state energy $E_0$ as a function of $1/L$ obtained from the DMRG. Other parameters are $J=1$, $U=10$, $\gamma=0.2$, $\alpha=1/3$, and $f=N/L=2/3$ [the same as those in Fig. 3(e,f)].}
%\label{figS3}
%\end{figure}

The TDVP method is similar to the DMRG implementation except the third step~\cite{PhysRevB.94.165116}, where eigensolver is replaced by the application of matrix exponential on the local wavefunction: $\tilde{M}^{j_n,j_{n+1}}=\mathrm{exp}(-i\hat{H}_\mathrm{eff}\tau){M}^{j_n,j_{n+1}}$ with a small time step $\tau$. There needs an addition operation in step 4 after the SVD, backward evolution of $\tilde{\Lambda}^{[n]}\tilde{B}^{j_{n+1}}$ (when sweeping from left) or $\tilde{A}^{j_n}\tilde{\Lambda}^{[n]}$ (when sweeping from right) before preparing the tensor $M^{j_{n+1},j_{n+2}}$ or $M^{j_{n-1},j_{n}}$ and this operation is actually similar to step 3 with time step $-\tau$ and a different effective Hamiltonian~\cite{PhysRevB.94.165116}. By sweeping from $n=1$ to $n=L-1$, the initial wavefunction $\ket{\Psi(0)}$ is involved to $\ket{\Psi(\tau)}$, and long time involved wavefunction can be obtained by repeated sweeps along the lattice. In the non-Hermitian version, the wavefunction after each sweep needs normalization because the evolution operator $\mathrm{exp}(-i\hat{H}_\mathrm{eff}\tau)$ is not unitary. In practice, we employ the Krylov-subspace approach in Arnoldi formalism~\cite{S0036142995280572} and combine step 2 and 3 in a matrix free style to apply the non-Hermitian effective Hamiltonian matrix exponential on the local wavefunction. Figure.~\ref{figS2} (c) shows comparison of numerical results of the expected occupation $n_{c}(t)=\braket{\Psi^r(t)|\hat{n}_{c}|\Psi^r(t)}$ obtained from the ED and TDVP methods with $c=11$ as the central site of a $L=21$ lattice. The initial state $\ket{\Psi^r(0)}$ is prepared in a product state where $N=3$ bosons locate in the center of the lattice. The relaxation of $n_{c}$ obtained from the ED (asterisk points) and TDVP (solid curves) methods consistent with each other for all parameters simulated. The system for $J_r=J$ and $J_l=e^{i\gamma}J$ with complex energy spectrum is also simulated and displayed in Fig.~\ref{figS2} (f).

Our implementation of the non-Hermitian DMRG and TDVP algorithms is mainly based on the ITensor library \cite{ITensor} and the demonstration codes can be found at \url{https://github.com/PhyRespo/ITensor}.

\end{appendix}

\end{acknowledgments}

\bibliography{reference}{}
\bibliographystyle{apsrev4-1}

\end{document}